\batchmode
\documentstyle[12pt]{article}
\newcommand{\nc}{\newcommand}
\nc{\renc}{\renewcommand}

\nc{\half}{{\textstyle{1\over2}}}
\nc{\etal}{\mbox{\it et al. }}
\nc{\ie}{{\it i.e.}}
\nc{\eg}{{\it e.g.}}

\renc{\thefootnote}{\arabic{footnote}}
\nc{\capt}[1]{{\bf Figure.} {\small\sl #1}}

\nc{\eqs}[2]{\mbox{Eqs.~(\ref{#1},\,\ref{#2})}}
\nc{\eq}[1]{\mbox{Eq.~(\ref{#1})}}

\nc{\figs}[2]{\mbox{Figs.~(\ref{#1},\,\ref{#2})}}
\nc{\fig}[1]{\mbox{Fig~.(\ref{#1})}}

\nc{\tag}[1]{\label{#1} \marginpar{{\footnotesize #1}}}
\nc{\mtag}[1]{\label{#1} \mbox{\marginpar{{\footnotesize #1}}}}
\renc{\baselinestretch}{1.5}
\newlength{\overeqskip}
\newlength{\undereqskip}
\nc{\be}[1]{\begin{equation} \mbox{$\label{#1}$}}
\nc{\bea}[1]{\begin{eqnarray} \mbox{$\label{#1}$}}
\nc{\Section}[2]{\section{#2}\label{#1}}
\nc{\Bibitem}[1]{\bibitem{#1}}
\nc{\Label}[1]{\label{#1}}

\nc{\eea}{\vspace{\undereqskip}\end{eqnarray}}
\nc{\ee}{\vspace{\undereqskip}\end{equation}}
\nc{\bdm}{\begin{displaymath}}
\nc{\edm}{\end{displaymath}}
\nc{\dpsty}{\displaystyle}
\nc{\bc}{\begin{center}}
\nc{\ec}{\end{center}}
\nc{\ba}{\begin{array}}
\nc{\ea}{\end{array}}
\nc{\bab}{\begin{abstract}}
\nc{\eab}{\end{abstract}}
\nc{\btab}{\begin{tabular}}
\nc{\etab}{\end{tabular}}
\nc{\bit}{\begin{itemize}}
\nc{\eit}{\end{itemize}}
\nc{\ben}{\begin{enumerate}}
\nc{\een}{\end{enumerate}}
\nc{\bfig}{\begin{figure}}
\nc{\efig}{\end{figure}}
\nc{\arreq}{&\!=\!&}
\nc{\arrmi}{&\!-\!&}
\nc{\arrpl}{&\!+\!&}
\nc{\arrap}{&\!\!\!\approx\!\!\!&}
\nc{\non}{\nonumber\\*}
\nc{\align}{\!\!\!\!\!\!\!\!&&}

\def\lsim{\; \raise0.3ex\hbox{$<$\kern-0.75em
      \raise-1.1ex\hbox{$\sim$}}\; }
\def\gsim{\; \raise0.3ex\hbox{$>$\kern-0.75em
      \raise-1.1ex\hbox{$\sim$}}\; }
\nc{\DOT}{\hspace{-0.08in}{\bf .}\hspace{0.1in}}
\nc{\Laada}{\hbox {$\sqcap$ \kern -1em $\sqcup$}}
\nc\loota{{\scriptstyle\sqcap\kern-0.55em\hbox{$\scriptstyle\sqcup$}}}
\nc\Loota{{\sqcap\kern-0.65em\hbox{$\sqcup$}}}
\nc\laada{\Loota}
\nc{\qed}{\hskip 3em \hbox{\BOX} \vskip 2ex}

\nc{\real}{{\rm I \! R}}
\nc{\Z}{{\sf Z \!\!\! Z}}
\nc{\complex}{{\rm C\!\!\! {\sf I}\,\,}}
\def\bigid{\leavevmode\hbox{\small1\kern-3.8pt\normalsize1}}
\def\id{\leavevmode\hbox{\small1\kern-3.3pt\normalsize1}}
\nc{\slask}{\!\!\!/}
\nc{\bis}{{\prime\prime}}
\nc{\pa}{\partial}
\nc{\na}{\nabla}
\nc{\ra}{\rangle}
\nc{\la}{\langle}
\nc{\goto}{\rightarrow}
\nc{\swap}{\leftrightarrow}

\nc{\EE}[1]{ \mbox{$\cdot10^{#1}$} }
\nc{\abs}[1]{\left|#1\right|}
\nc{\at}[2]{\left.#1\right|_{#2}}
\nc{\norm}[1]{\|#1\|}
\nc{\abscut}[2]{\Abs{#1}_{\scriptscriptstyle#2}}
\nc{\vek}[1]{{\rm\bf #1}}
\nc{\integral}[2]{\int\limits_{#1}^{#2}}
\nc{\inv}[1]{\frac{1}{#1}}
\nc{\dd}[2]{{{\partial #1}\over{\partial #2}}}
\nc{\ddd}[2]{{{{\partial}^2 #1}\over{\partial {#2}^2}}}
\nc{\dddd}[3]{{{{\partial}^2 #1}\over
	{\partial #2 \partial #3}}}
\nc{\dder}[2]{{{d #1}\over{d #2}}}
\nc{\ddder}[2]{{{d^2 #1}\over{d {#2}^2}}}
\nc{\dddder}[3]{{d^2 #1}\over
	{d #2 d #3}}
\nc{\dx}[1]{d\,^{#1}x}
\nc{\dy}[1]{d\,^{#1}y}
\nc{\dz}[1]{d\,^{#1}z}
\nc{\dl}[1]{\frac{d\,^{#1}l}{(2\pi)^{#1}}}
\nc{\dk}[1]{\frac{d\,^{#1}k}{(2\pi)^{#1}}}
\nc{\dq}[1]{\frac{d\,^{#1}q}{(2\pi)^{#1}}}

\nc{\cc}{\mbox{$c.c.$ }}
\nc{\hc}{\mbox{$h.c.$ }}
\nc{\cf}{cf.\ }
\nc{\erfc}{{\rm erfc}}
\nc{\Tr}{{\rm Tr\,}}
\nc{\tr}{{\rm tr\,}}
\nc{\pol}{{\rm pol}}
\nc{\sign}{{\rm sign}}
\nc{\bfT}{{\bf T }}

\def\GeV{{\rm\ GeV}}

\nc{\cA}{{\cal A}}
\nc{\cB}{{\cal B}}
\nc{\cD}{{\cal D}}
\nc{\cE}{{\cal E}}
\nc{\cG}{{\cal G}}
\nc{\cH}{{\cal H}}
\nc{\cL}{{\cal L}}
\nc{\cO}{{\cal O}}
\nc{\cT}{{\cal T}}
\nc{\cN}{{\cal N}}
\nc{\rvac}[1]{|{\cal O}#1\rangle}
\nc{\lvac}[1]{\langle{\cal O}#1|}
\nc{\rvacb}[1]{|{\cal O}_\beta #1\rangle}
\nc{\lvacb}[1]{\langle{\cal O}_\beta #1 |}
\nc{\bb}{\bar{\beta}}
\nc{\bt}{\tilde{\beta}}
\nc{\ctH}{\tilde{\cal H}}
\nc{\chH}{\hat{\cal H}}
\nc{\1}{\aa}
\nc{\2}{\"{a}}
\nc{\3}{\"{o}}
\nc{\4}{\AA}
\nc{\5}{\"{A}}
\nc{\6}{\"{O}}
\nc{\al}{\alpha}
\nc{\g}{\gamma}
\nc{\Del}{\Delta}
\nc{\e}{\epsilon}
\nc{\eps}{\epsilon}
\nc{\lam}{\lambda}
\nc{\om}{\omega}
\nc{\Om}{\Omega}
\nc{\ve}{\varepsilon}
\nc{\mn}{{\mu\nu}}
\nc{\k}{\kappa}
\nc{\vp}{\varphi}

\nc{\advp}[3]{{\it  Adv.\ in\ Phys.\ }{{\bf #1} {(#2)} {#3}}}
\nc{\annp}[3]{{\it  Ann.\ Phys.\ (N.Y.)\ }{{\bf #1} {(#2)} {#3}}}
\nc{\apl}[3]{{\it  Appl. Phys. Lett. }{{\bf #1} {(#2)} {#3}}}
\nc{\apj}[3]{{\it  Ap.\ J.\ }{{\bf #1} {(#2)} {#3}}}
\nc{\apjl}[3]{{\it  Ap.\ J.\ Lett.\ }{{\bf #1} {(#2)} {#3}}}
\nc{\app}[3]{{\it Astropart.\ Phys.\ }{{\bf #1} {(#2)} {#3}}}
\nc{\cmp}[3]{{\it  Comm.\ Math.\ Phys.\ }{{ \bf #1} {(#2)} {#3}}}
\nc{\cqg}[3]{{\it  Class.\ Quant.\ Grav.\ }{{\bf #1} {(#2)} {#3}}}
\nc{\epl}[3]{{\it  Europhys.\ Lett.\ }{{\bf #1} {(#2)} {#3}}}
\nc{\ijmp}[3]{{\it Int.\ J.\ Mod.\ Phys.\ }{{\bf #1} {(#2)} {#3}}}
\nc{\ijtp}[3]{{\it Int.\ J.\ Theor.\ Phys.\ }{{\bf #1} {(#2)} {#3}}}
\nc{\jmp}[3]{{\it  J.\ Math.\ Phys.\ }{{ \bf #1} {(#2)} {#3}}}
\nc{\jpa}[3]{{\it  J.\ Phys.\ A\ }{{\bf #1} {(#2)} {#3}}}
\nc{\jpc}[3]{{\it  J.\ Phys.\ C\ }{{\bf #1} {(#2)} {#3}}}
\nc{\jap}[3]{{\it J.\ Appl.\ Phys.\ }{{\bf #1} {(#2)} {#3}}}
\nc{\jpsj}[3]{{\it J.\ Phys.\ Soc.\ Japan\ }{{\bf #1} {(#2)} {#3}}}
\nc{\lmp}[3]{{\it Lett.\ Math.\ Phys.\ }{{\bf #1} {(#2)} {#3}}}
\nc{\mpl}[3]{{\it  Mod.\ Phys.\ Lett.\ }{{\bf #1} {(#2)} {#3}}}
\nc{\ncim}[3]{{\it  Nuov.\ Cim.\ }{{\bf #1} {(#2)} {#3}}}
\nc{\np}[3]{{\it  Nucl.\ Phys.\ }{{\bf #1} {(#2)} {#3}}}
\nc{\npps}[3]{{\it  Nucl.\ Phys.\ Proc.\ Suppl.\ }{{\bf #1} {(#2)} {#3}}}
\nc{\pr}[3]{{\it Phys.\ Rev.\ }{{\bf #1} {(#2)} {#3}}}
\nc{\pra}[3]{{\it  Phys.\ Rev.\ A\ }{{\bf #1} {(#2)} {#3}}}
\nc{\prb}[3]{{\it  Phys.\ Rev.\ B\ }{{{\bf #1} {(#2)} {#3}}}}
\nc{\prc}[3]{{\it  Phys.\ Rev.\ C\ }{{\bf #1} {(#2)} {#3}}}
\nc{\prd}[3]{{\it  Phys.\ Rev.\ D\ }{{\bf #1} {(#2)} {#3}}}
\nc{\prl}[3]{{\it Phys.\ Rev.\ Lett.\ }{{\bf #1} {(#2)} {#3}}}
\nc{\pl}[3]{{\it  Phys.\ Lett.\ }{{\bf #1} {(#2)} {#3}}}
\nc{\prep}[3]{{\it Phys.\ Rep.\ }{{\bf #1} {(#2)} {#3}}}
\nc{\prsl}[3]{{\it Proc.\ R.\ Soc.\ London\ }{{\bf #1} {(#2)} {#3}}}
\nc{\ptp}[3]{{\it  Prog.\ Theor.\ Phys.\ }{{\bf #1} {(#2)} {#3}}}
\nc{\ptps}[3]{{\it  Prog\ Theor.\ Phys.\ suppl.\ }{{\bf #1} {(#2)} {#3}}}
\nc{\physa}[3]{{\it  Physica\ A\ }{{\bf #1} {(#2)} {#3}}}
\nc{\physb}[3]{{\it  Physica\ B\ }{{\bf #1} {(#2)} {#3}}}
\nc{\phys}[3]{{\it Physica\ }{{\bf #1} {(#2)} {#3}}}
\nc{\rmp}[3]{{\it  Rev.\ Mod.\ Phys.\ }{{\bf #1} {(#2)} {#3}}}
\nc{\rpp}[3]{{\it Rep.\ Prog.\ Phys.\ }{{\bf #1} {(#2)} {#3}}}
\nc{\sjnp}[3]{{\it Sov.\ J.\ Nucl.\ Phys.\ }{{\bf #1} {(#2)} {#3}}}
\nc{\spjetp}[3]{{\it Sov.\ Phys.\ JETP\ }{{\bf #1} {(#2)} {#3}}}
\nc{\yf}[3]{{\it Yad.\ Fiz.\ }{{\bf #1} {(#2)} {#3}}}
\nc{\zetp}[3]{{\it Zh.\ Eksp.\ Teor.\ Fiz.\  }{{\bf #1}  {(#2)} {#3}}}
\nc{\zp}[3]{{\it Z.\ Phys.\ }{{\bf #1} {(#2)} {#3}}}
\nc{\ibid}[3]{{\sl ibid.\ }{{\bf #1} {#2} {#3}}}
\nc{\rf}[1]{(\ref{#1})}
\nc{\nn}{\nonumber \\*}
\nc{\bfB}{\bf{B}}
\nc{\bfv}{\bf{v}}
\nc{\bfx}{\bf{x}}
\nc{\bfy}{\bf{y}}
\nc{\vx}{\vec{x}}
\nc{\vy}{\vec{y}}
\nc{\oB}{\overline{B}}
\nc{\oI}{\overline{I}}
\nc{\oR}{\overline{R}}
\nc{\rar}{\rightarrow}
\nc{\ti}{\times}
\nc{\slsh}{\hskip-5pt/}
\nc{\sm}{Standard~Model~}
\nc{\MP}{M_{\rm Pl}}
\nc{\tp}{t_{\rm Pl}}
\nc{\ave}{\bar{E}}

\nc{\eff}{{\rm eff}}
\nc{\kk}{\vek{k}}
\nc{\pp}{{\rm p}}
\nc{\ga}{g_{a\gamma}}
\nc{\vv}{\\}
\nc{\eee}{{\bf E}}
\nc{\bbb}{{\bf B}}
\nc{\qcd}{T_{\rm QCD}}
\nc{\G}{\rm \ G}
\def\vec#1{{\bf #1}}

\def\lae{\;^{<}_{\sim} \;}  

\begin{document}
{\title{\vskip-2truecm{\hfill {{\small \\
	\hfill HIP-1998-XX/th \\
	}}\vskip 1truecm}
{\bf MSSM Dark Matter Constraints and Decaying B-balls.}}
{\author{
{\sc  Kari Enqvist$^{1}$}\\
{\sl\small Department of Physics and Helsinki Institute of Physics,}\\ 
{\sl\small P.O. Box 9,
FIN-00014 University of Helsinki,
Finland}\\
{\sc and}\\
{\sc  John McDonald$^{2}$}\\
{\sl\small Department of Physics, P.O. Box 9,
FIN-00014 University of Helsinki,
Finland}
}
\maketitle
\begin{abstract}
\noindent

              In the MSSM, LSP dark matter could arise from the decay of B-balls rather than from 
thermal relics, with a quite different dependence on the MSSM parameters
and a natural correlation with the baryon asymmetry. We discuss the
 constraints imposed on the properties of B-balls 
and the MSSM spectrum by experimental constraints and show 
that B-balls cannot form from an Affleck-Dine condensate 
with 100$\%$ efficiency in the absence of 
LSP annihilations after they decay. For likely formation efficiencies the LSP and 
slepton masses are typically constrained to be light. The effects of LSP annihilations 
after the B-balls decay are discussed; for sufficiently small decay temperatures 
annihilations will play no role, opening up the possibility of experimentally testing the scenario.

\end{abstract}
\vfil
\footnoterule
{\small $^1$enqvist@pcu.helsinki.fi};
{\small $^2$mcdonald@rock.helsinki.fi}

\thispagestyle{empty}
\newpage
\setcounter{page}{1}


               The conventional view of the cosmology of the Minimal Supersymmetric Standard Model (MSSM) \cite{nilles} at temperatures less than that of the electroweak phase transition, $T_{ew}$, is of a radiation dominated Universe 
in which the baryon number is 
generated at temperatures greater than or equal to $T_{ew}$ and the 
dark matter is composed of thermal relic neutralino LSPs \cite{eu,susydm}. Recently it has been shown that the MSSM can sustain a very different post-inflation cosmology \cite{bbb1,bbb2}. In this scenario
the baryon asymmetry comes from a variant of the Affleck-Dine (AD) mechanism \cite{ad}, B-ball Baryogenesis (BBB) \cite{bbb1,bbb2}, in which the baryon number originates from the collapse of a d $\geq 6$ AD condensate 
to a mixture of free squarks and 
non-topological solitons, B-balls.
The reheating temperature must be less than $10^{3-5} \GeV$ in order that the B-balls 
do not thermalize \cite{bbb2}. For the particular case of a d=6 AD condensate, which we will focus on in the 
following, the Universe is dominated by the energy density of an inflaton down to temperatures typically of the order of 1 GeV, which is fixed by the observed baryon asymmetry when the CP violating phase responsible for the asymmetry is of the order
of 1 \cite{bbb2}. In particular, this is expected to be true for D-term inflation models \cite{kmr,bbbd}.
The resulting B-balls have large charges and typically decay at temperatures between 1 MeV and 1 GeV \cite{bbb2,bbbd}. Being made of squarks, they initially decay to LSPs and baryons with a similar number density. 
If the B-ball decay temperature is below the freeze-out temperature of the LSPs and
there are no subsequent annihilations of the LSPs, the similarity of the number densities will 
be preserved. In general, the resulting dark matter density will differ from that expected purely
 from thermal relics and will be determined by the MSSM parameters together with the B-ball parameters.
 Given the B-ball parameters, the MSSM parameters can be constrained by the resulting
 density of dark matter. Conversely, the B-ball parameters can be
 constrained by the dark matter density for a given set of MSSM parameters. 
Such constraints will serve as an important test of the scenario 
once the B-ball parameters, which are calculable (although non-trivial), are better known. 

          Under the assumption that $\Omega = 1$ as a result of a period of inflation, primordial nucleosynthesis bounds on the density of baryons in the Universe \cite{sarkar} suggest that 90$\%$ of the mass in the Universe must be in the form of 
non-baryonic dark matter, a conclusion supported by structure formation models and by direct observations of galactic rotation curves \cite{eu}. (One should, however, note that recent supernova surveys \cite{SN} indicate 
that there might exist a non-zero cosmological
constant $\Lambda$. If so, the matter density could be much less than the critical density).
Recent observations of
deuterium features in the spectra of quasars \cite{tytler}, as well as 
determinations of the metallicity of extragalactic HII regions 
on which extrapolations of the primordial $^{4}$He abundance are based, 
have led to a somewhat conflicting picture of the baryon asymmetry \cite{gary}.
The baryon asymmetry appears to be either relatively high, as favoured by 
deuterium observations, in which case there might be
a problem with the $^{4}$He abundance, or relatively low, in which case 
there might a be problem understanding the stellar evolution of 
 the high
D/$^3$H abundance required. Because of this unclear situation, which 
is likely to be due to hidden systematic errors, 
for the purposes of the present paper we will adopt 
a conservative nucleosynthesis bound  \cite{sarkar} on the baryon density,
$0.0048 \lae \Omega_{B}h^{2} \lae 0.013$, where the age of the Universe 
requires that $h$ satisfies 
$0.4 \lae h \lae 0.65$ for an $\Omega = 1$ Universe. 
The ratio of the number density of baryons to dark matter particles, 
$\sigma_{B}$, is then constrained to satisfy
\be{sigma}  \sigma_{B} = (0.48-1.3) 
\times 10^{-2} h^{-2} \frac{m_{DM}}{m_{N}}    ~,
\ee
where $m_{DM}$ is the mass of the dark matter particle and $m_{N}$ is 
the nucleon mass.
For example, for the case of dark matter particles with weak scale masses, with $h \approx 0.5$ 
and $m_{DM} \approx m_{W}$ we obtain $\sigma_{B}\approx 1.5-4$; a highly suggestive result. 
Most baryogenesis mechanisms give no explicit connection between the densities of 
baryons and dark matter; it is implicitly assumed that their similarity is a result either of 
coincidence or of some hidden anthropic selection effect \cite{as}, neither of which is particularly satisfying as an explaination.
 The similarity of the dark matter and baryon densities (in particular their number densities) can best be explained if they are produced by the 
same mechanism\footnote{\small For an alternative connection between 
B-balls and the baryon to dark matter ratio, 
in the context of gauge-mediated SUSY breaking models with stable B-balls \cite{stableQ}, see reference \cite{ls}.}.

             If the reheating temperature is much less than $T_{f}$, there will be essentially no
thermal relic background of LSPs, since the additional entropy released during the inflaton matter domination period will strongly suppress the thermal relic density by a factor $(T_{R}/T_{f})^{5}$. (The LSPs have a freeze-out temperature of
$T_{f} \approx m_{{\rm LSP}}/20$ \cite{susydm,jdm1}, where $m_{{\rm LSP}}$ is the LSP neutralino mass. The present direct experimental bound on the LSP mass, valid for any
$\tan\beta$ (but assuming $m_{\tilde\nu}\ge 200$ GeV),
is $m_{{\rm LSP}}\ge 25$ GeV \cite{lspbound}.
If one assumes the MSSM with universal soft SUSY breaking masses and unification \cite{nilles}, LEP results
can be combined to yield an excluded region in the 
$(m_{{\rm LSP}}, m_{\tilde l_R})$-plane \cite{lep1}. In the case of $\tilde e_R$, which
provides the most stringent bound, the excluded region is roughly
parametrized by $m_{{\rm LSP}}\lsim 0.95 m_{\tilde e_R}$ for $45\GeV\lsim
m_{\tilde e_R}
\lsim 78$ GeV (this result holds for ${\rm tan}\beta=2$ and $\mu=-200$
GeV) \cite{lep1}. Therefore the LSP freeze-out 
temperature is expected to be greater than about 1-2 GeV).
Thus there are two possibilities, depending on 
$T_R$ and $T_f$: either the
LSP cold dark matter density, 
$\Omega_{\rm LSP}$, will be given
solely by the  LSP density which originated from the B-ball decay, which
we denote by $\Omega_{\rm BB}$,
or there will also be a relic density so that $\Omega_{\rm LSP}=\Omega_{\rm BB}
+\Omega_{\rm relic}$. We will consider both possibilities in the following. 

              Let us first discuss some general aspects of the production of 
neutralino dark matter via B-ball decays. 
Once the d=6 AD condensate collapses, a fraction $f_{B}$ of the total 
B asymmetry ends up in the form of B-balls.
The B-balls have charges $B \approx 10^{23}f_{B}(1\GeV/T_{R})$ \cite{bbb2}
and subsequently decay at a temperature
\be{decayT} T_{d} \approx 0.01 
\left(\frac{f_{s}}{f_{B}}\right)^{1/2}
\left(\frac{0.01}{|K|}\right)^{3/4}
\left(\frac{m}{100 \GeV}\right)
\left(\frac{T_R}{1 \GeV}\right)^{1/2}
\GeV
~,\ee
where $m$ is the B-ball squark mass and $f_{s}$ is the possible enhancement factor if the squarks can decay to a pair of scalars rather
than to final states with two fermions; we have estimated $f_{s} \approx 10^{3}$ \cite{bbb2}. 
($f_{B}$ and $T_{d}$ are the only B-ball parameters which enter into the determination of the LSP density from B-ball decay). 
For example, with $T_{R} \approx 1 \GeV$, as suggested by the d=6 AD mechanism, 
and with $f_{B}$ in the range 0.1 to 1 (in accordance with our previous argument \cite{bbb2} 
that B-ball formation from an AD condensate is likely to be 
very efficient, although the numerical value of $f_{B}$ is not yet known), $T_{d}$ will generally be in the range 1 MeV to 1 GeV. 
In this case the 
B-balls will decay below the LSP freeze-out temperature.
The neutralino density will then consist of a possible thermal relic component, $n_{relic}(T)$, and a 
component from B-ball decays, $n_{BB}(T)$. The value of $n_{BB}(T)$ will depend upon whether or not the 
LSPs from B-ball decay can subsequently annihilate. The upper limit on $n_{LSP}(T)$ from 
annihilations is given by
\be{ann} n_{LSP}(T) \lae n_{limit}(T) \equiv \left( \frac{H}{<\sigma v>_{ann}} \right)_{T}  ~,\ee
where $<\sigma v>_{ann}$ is the thermal average of the annihilation cross-section times the 
relative velocity of the LSPs, which can be generally written in the form
$<\sigma v>_{ann} = a + b T / m_{{\rm LSP}}$ \cite{susydm}.
If $n_{LSP}(T) \lae n_{limit}(T)$, and if $f_{B}$ is not too small compared with 1, 
there will be a natural similarity between the number density of LSPs and that of the baryons. 
Otherwise the annihilation of neutralinos will suppress the number density of LSPs relative to that 
of the baryons, although we will still have an interesting non-thermal neutralino relic density.

             Assuming that $n_{LSP}(T_{d}) \lae n_{limit}(T_{d})$, the LSP density from B-ball
 decays will be given by
 \be{nd} n_{BB} = 3 f_{B} n_{B}     ~,\ee
where three LSPs are produced per unit baryon number from the decay of the B-ball squarks.
Thus the B-ball produced LSP and baryonic densities will be related by
\be{bdm} \frac{\Omega_{B}}{\Omega_{BB}} 
= {m_N \over 3 f_{B}m_{\rm LSP}} ~.
\ee
B generation via the AD mechanism requires inflation \cite{ad,kmr}, and although varieties of inflationary
models exist with $\Omega_{\rm tot}<1$, we will nevertheless adopt the point of view that
inflation implies $\Omega_{\rm tot}=1$ to a high precision. We may then write
\bea{omega1}
\Omega_{\rm tot}&=&\Omega_0+\Omega_{\rm LSP}+\Omega_B\nn
&=&\Omega_0+\Omega_{\rm relic}+\left({3 f_{B}m_{\rm LSP}\over m_N}
+1\right) \Omega_B =1~,
\eea
where $\Omega_0$ includes the hot dark matter (HDM) 
component and/or a possible cosmological constant.
Therefore $\Omega_{B}$ is fixed by $\Omega_0$,  $f_{B}$ 
and $m_{{\rm LSP}}$ together with the MSSM parameters entering into the annihilation rate. Applying nucleosynthesis bounds on
$\Omega_{B}$ then gives constraints on these parameters. Note that, so long as LSP annihilations 
after B-ball decay can be neglected, the resulting LSP density is independent of $T_{d}$.

        Let us first consider the case where the thermal relic density $\Omega_{relic}$
is negligible. This would be true if $T_R$ was sufficiently small compared with the freeze-out temperature $T_f$.  We then obtain 
the limit
\be{b1} 76.9 (1-\Omega_0) h^{2} - 1 \lsim \frac{3 m_{{\rm LSP}}f_{B}}{m_{N}} 
 \lsim  208.3 (1-\Omega_0) h^{2} - 1~.\ee
With $\Omega_0 = 0$ this would result in a bound on the LSP mass given by
\be{b2} 
3.8 
f_{B}^{-1} \GeV 
\lsim m_{{\rm LSP}}
\lsim 
29 
 f_{B}^{-1} \GeV  ~,
\ee
where we have used $0.4 \lae h \lae 0.65$.
If $f_{B} = 1$ this would be only marginally compatible with present experimental constraints and then only if we do not consider universal soft SUSY breaking masses.
Larger values of $\Omega_0$ impose even tighter bounds on $m_{{\rm LSP}}$, requiring $f_{B} 
< 1$. Therefore, in the absence of annihilations after 
B-ball decays, LSP dark matter from B-balls is likely to be compatible with nucleosynthesis bounds only if a significant fraction of the baryon asymmetry exists outside the B-balls. 
Reasonable values of $f_{B}$ can, however, accomodate an interesting range of LSP masses; for 
example, values in the range 0.1 to 1 (which we consider to be reasonable) allow LSP masses as large as 290 GeV. $f_{B}$ can be calculated theoretically, but this requires an analysis of the non-linear evolution of the unstable AD condensate \cite{emn}.
The comparison of the theoretical value with the dark matter constraints 
will be an important test of this scenario. 

\begin{figure}
\leavevmode
\centering
\vspace*{95mm} 
\includegraphics{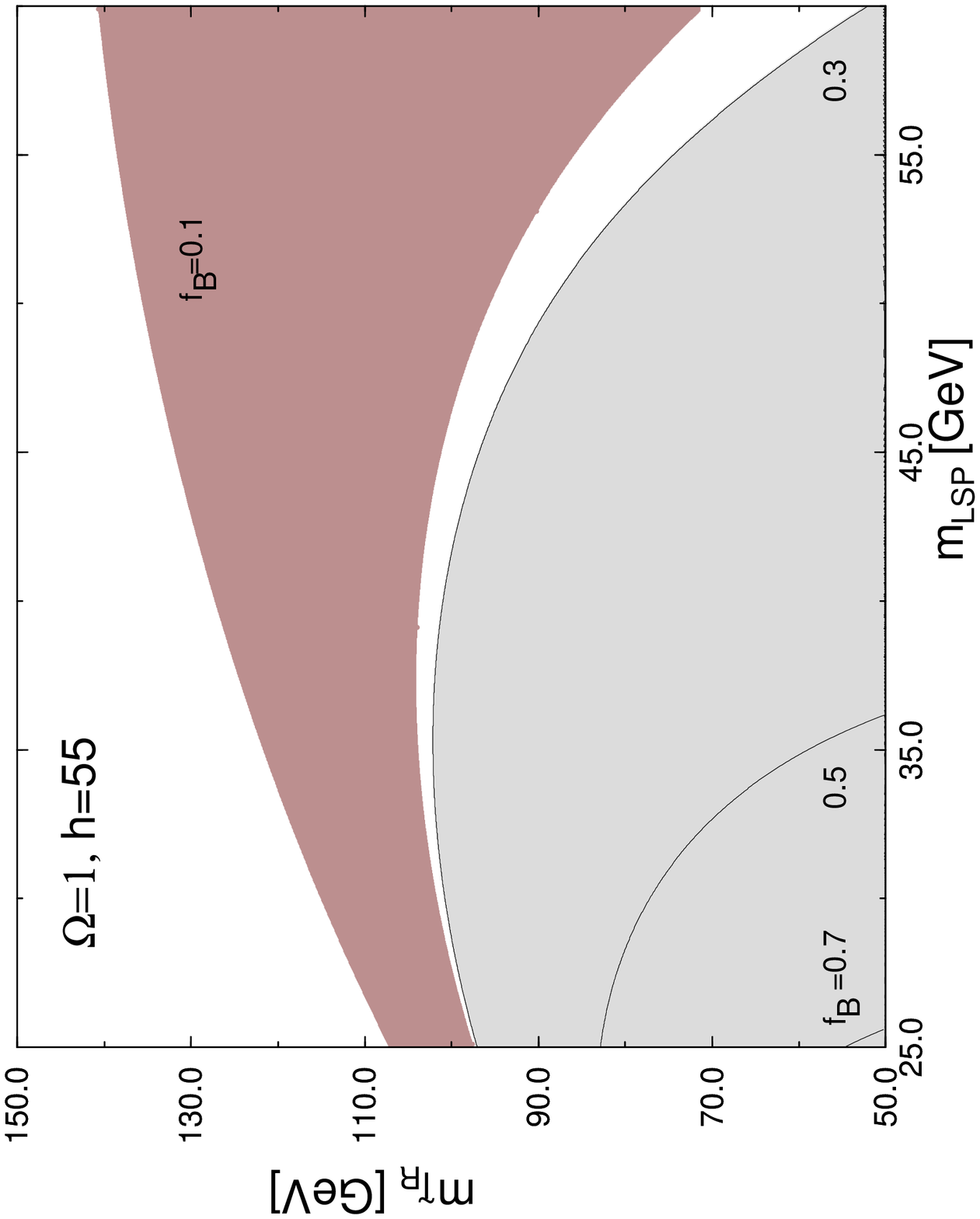}
\includegraphics{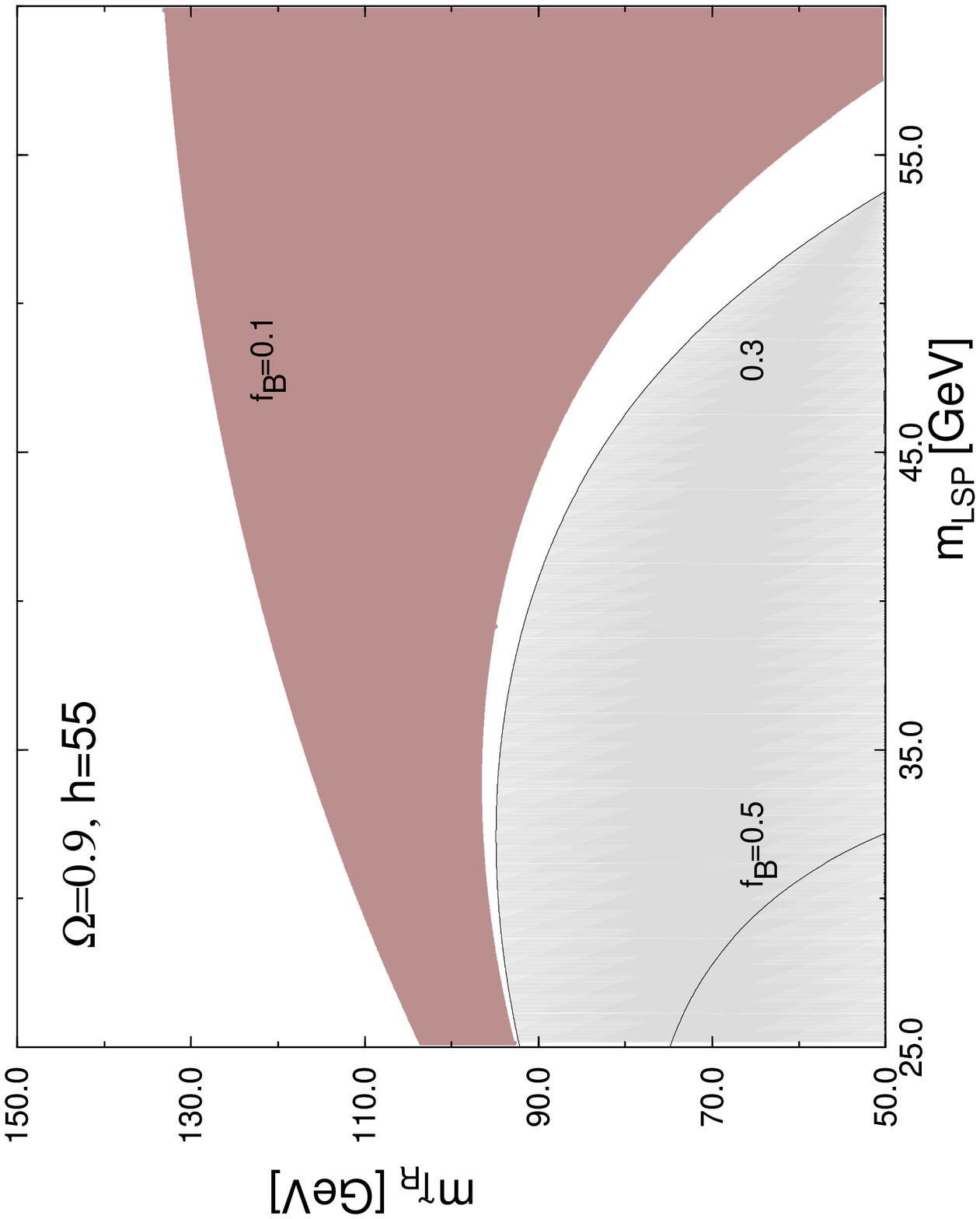}   
\includegraphics{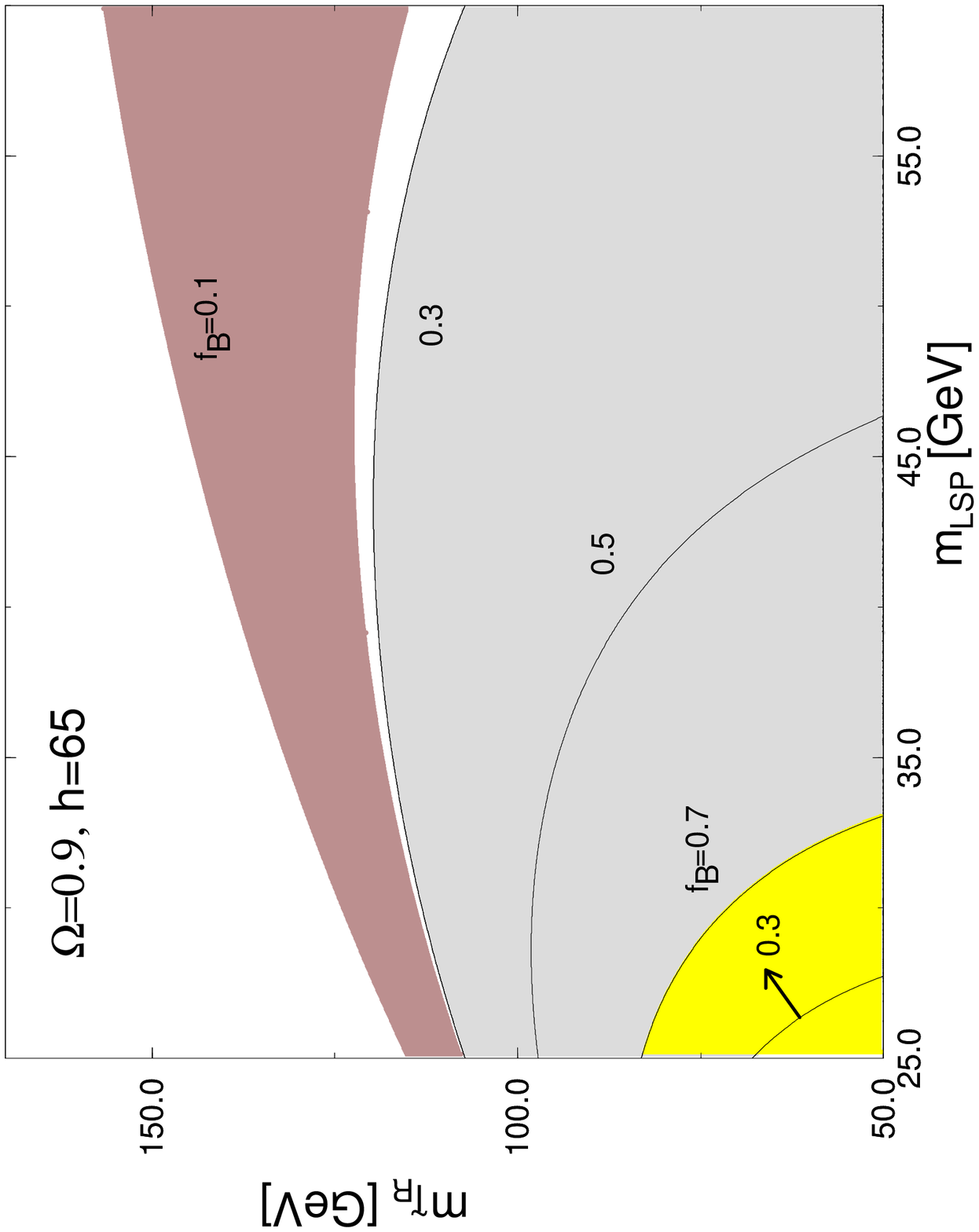}
\includegraphics{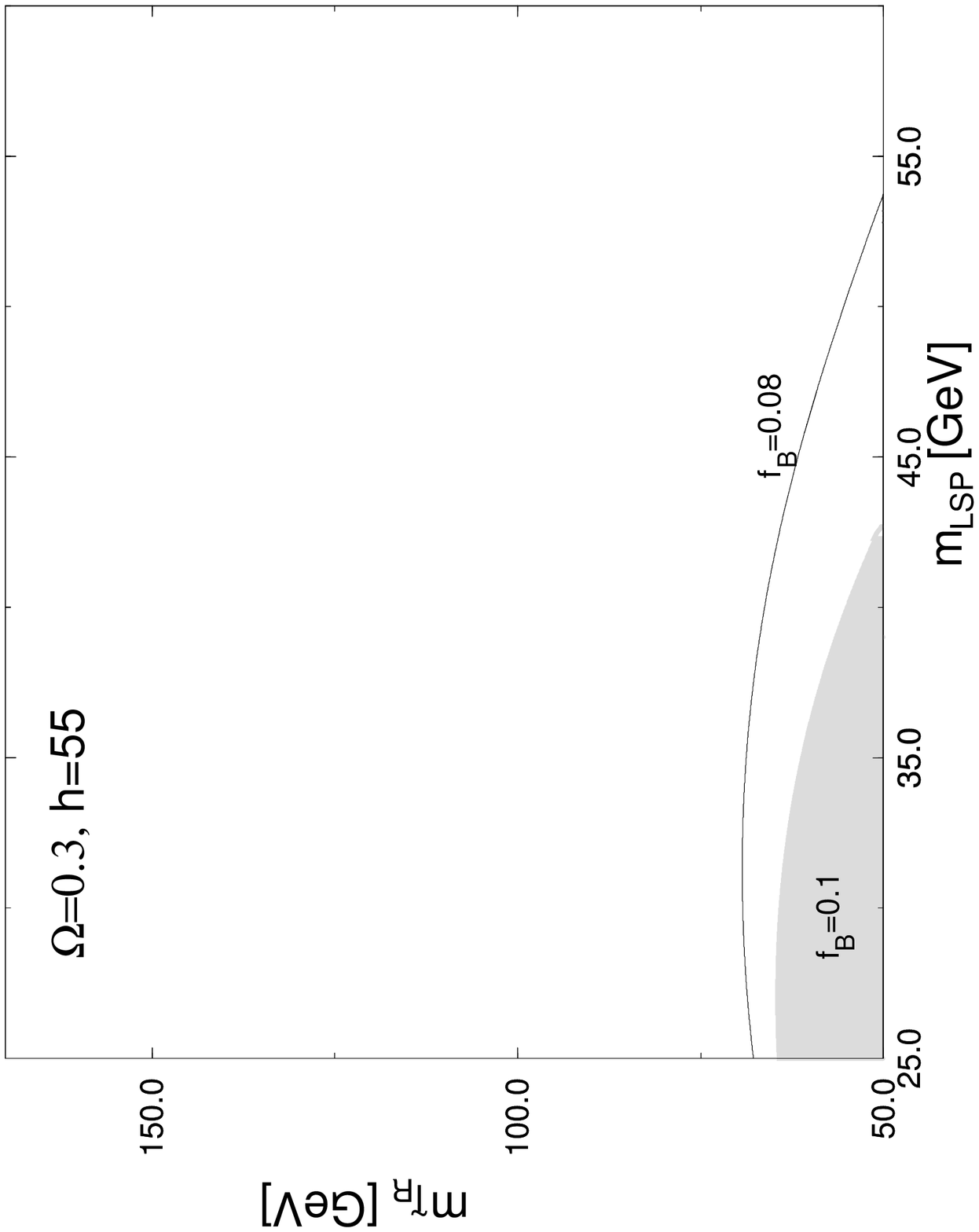}
\caption{The allowed regions in the $(m_{\rm LSP},m_{\tilde l_R})$-plane
for different values of the total CDM density $\Omega$ and 
the Hubble parameter $h$.}
\label{kuva1}       
\end{figure} 

       We next consider the case with $T_R>T_f$. In this case 
there will be a significant thermal relic density 
and we can use nucleosynthesis 
bounds on $\Omega_{B}$ to constrain the masses of the particles 
responsible for the LSP annihilation cross-section. 
The constraints will depend on the identity of the LSP and the masses of the particles 
entering the LSP annihilation cross-section. In general, this would require
 a numerical analysis of the renormalization group equations for the SUSY particle spectrum.
However, for the case of universal scalar and gaugino masses at a large scale, the LSP 
is likely to be mostly bino and the lightest scalars are likely to be the right-handed sleptons \cite{wells}. This is consistent with the requirement that the LSP does not have a large coupling to the Z boson, which would otherwise efficiently annihilate away the thermal relics. However, there will be a small, model-dependent Higgsino component which will be important for LSP masses close to the Z pole. For LSP masses away from this pole, it will be a reasonable approximation
to treat the LSP as a $pure$ bino. In this preliminary study we will consider the pure bino approximation, although the possible suppression of the thermal relic density around the Z pole 
due to a Higgsino component and the subsequent weakening of MSSM constraints should be kept in mind. 

        For the case of a pure bino, the largest contribution to the annihilation cross-section 
comes from
$t$-channel $\tilde l_R$ exchange in ${\chi}{\rm \chi}\to l^+l^-$ \cite{bino}. 
In that case one finds \cite{wells}
\be{binos}
\Omega_{\rm relic}h^2={\Sigma^2\over M^2m_{\rm LSP}^2}
\left[\left(1-{m_{\rm LSP}^2\over \Sigma}\right)^2
+{m_{\rm LSP}^4 \over \Sigma^2}\right]^{-1}~,
\ee
where $M\approx 1$ TeV and $\Sigma=m_{\rm LSP}^2+m_{\tilde l_R}^2$. 
Plugging this into \eq{omega1}
and using the range of $\Omega_B$ allowed by nucleosynthesis, 
one may obtain allowed ranges
in the $(m_{\rm LSP},m_{\tilde l_R})$-plane. 
These are demonstrated in Fig. 1 for different values of
$\Omega_0$ and $h$. 

In the conventional MSSM case \eq{binos} would imply that both $m_{\rm LSP}$
and $m_{\tilde l_R}$ should be less than about 200 GeV. 
Because of the added B-ball contribution
a more stringent constraint follows in the present case. 
If the reheating temperature is
larger than the LSP freeze-out temperature, and if we consider the range $0.1 \lae f_B \lae 1$ to be the most likely, we may conclude that 
only a {\it very light
sparticle spectrum} is consistent with $\Omega = 1$; this is so in particular 
if there is a
cosmological constant with $\Omega_0\approx 0.7$, as suggested by recent 
supernova 
studies \cite{SN}. In any case, it is evident that in the case $T_R>T_f$ 
we obtain significant constraints on the B-ball formation efficiency from MSSM constraints. 
This is demonstrated in Fig. 2 for the case of 
$\Omega_0=0.1$,
where we plot the allowed regions in the $(f_B, m_{\rm LSP})$-plane for 
fixed values of $m_{\tilde l_R}$. 
\begin{figure}
\leavevmode
\centering
\vspace*{85mm} 
\includegraphics{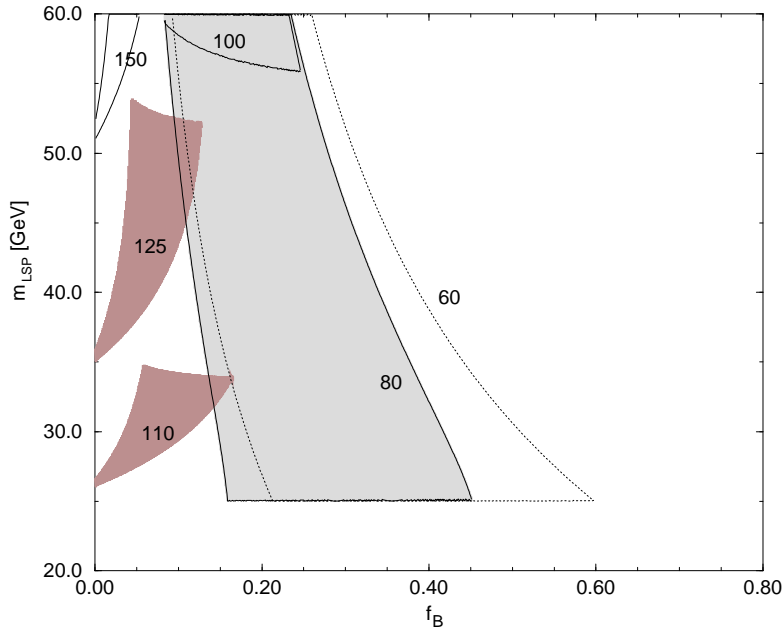}
\caption{The allowed region in the $(f_B, m_{\rm LSP})$-plane
for fixed $m_{\tilde l_R}$, assuming that
the total CDM density $\Omega=0.9$ and 
the Hubble parameter $h=0.65$.}
\label{kuva2}       
\end{figure} 
As can be seen, in the case of $T_R>T_f$ dark matter constrains $f_B$ to be 
less than about 0.6. If the SUGRA-based LEP limit 
 $m_{{\rm LSP}}\lsim 0.95 m_{\tilde e_R}$ ($45\GeV\lsim 
m_{\tilde e_R}\lsim 78$ GeV)
 is implemented \cite{lep1}, the limit on $f_B$
would be even lower. This serves to emphasize the need for an accurate
 theoretical determination of  
$f_B$.

            So far we have considered B-ball decay in the absence of LSP 
annihilations. However, depending on $T_{d}$, annihilations after the B-balls decay 
may significantly reduce the final LSP density. For the case where 
$T_{R} > T_{d}$, the limiting density for pure binos 
may be expressed in terms of the thermal relic density at present as 
\be{an1} n_{limit}(T_{d})  
\approx \frac{g(T_{f})}{g(T_{\gamma})} 
\frac{T_{f}^{2}\; T_{d}^{2} \rho_{o}}{T_{b} \; T_{\gamma}^{3} m_{LSP}}
\Omega_{relic} 
~,\ee
where $T_{b}$ is the temperature above which the $b$ term in the thermally averaged annihilation cross-section comes to dominate, $T_{\gamma} \approx 2.4 \times 10^{13}\GeV$ is the present photon temperature and $\rho_{o} = 7.5 \times 10^{-47} h^{2} \GeV^{4}$ is the present energy density 
of the Universe. Assuming that $T_{d}$ is sufficiently small compared with $T_{f}$, in order that the thermal relic density can be neglected compared with the limiting density, the condition for the annihilations to be negligible becomes
\be{an2} f_{B} \leq f_{B\;c} \approx 0.1
\left(\frac{5 \times 10^{-11}}{\eta_{B}}\right) 
\left(\frac{100\GeV}{m_{LSP}}\right) 
\left(\frac{1\GeV}{T_{d}}\right)^{2} 
\left(\frac{T_{d}}{T_{b}}\right) 
\ee
 $${\times \left(\frac{\Sigma^{1/2}}{100 \GeV}\right)^{4} 
\left[ \left(1- \frac{m_{LSP}^{2}}{\Sigma}\right)^{2} 
+ \frac{m_{LSP}^{4}}{\Sigma^{2}}  \right]^{-1}
},$$
where the baryon asymmetry is constrained by nucleosynthesis 
to be in the range $\eta_{B} = (3-8) \times 10^{-11}$. 
In this we have used $T_{f} \approx m_{LSP}/20$. 
If annihilations are significant, the 
LSP density is given by that expected from B-ball decays without annihilations but with 
$f_{B}$ replaced by $f_{B\:c}$. 
Therefore, so long as $f_{B\;c}$ is not very much smaller than 1, there will still be a similar number density of baryons and dark matter in this case.
If $T_{d} > T_{R}$, the B-balls will decay during the inflaton matter dominated era and the 
baryon number and limiting 
density will differ from the case where B-balls decay during radiation domination. This 
results in a stronger bound on $f_{B}$, 
\be{an4} f_{B} \lae \frac{2}{5} \left(\frac{T_{R}}{T_{d}}\right)^{3} f_{B\;c} \;\;,\;\;\;\; T_{d} > T_{R}    ~.\ee
Thus values of $T_{R}$ less than $T_{d}$ are strongly disfavoured.

            For the case of a pure bino, the $a$ and $b$ entering the annihilation cross-section are given by \cite{susydm}
\be{an5} a = \frac{1}{2 \pi} \frac{p}{m_{LSP}} \left( \frac{g_{1}^{2}
 }{2 m_{\tilde{l}_{R}}^{2}}\right)^{2} m_{\tau}^{2} \; ; \;\;\;\;\;
b = \frac{6}{\pi} \frac{p}{m_{LSP}} \left( \frac{g_{1}^{2}}{ 2 m_{\tilde{l}_{R}}^{2}}
\right)^{2} m_{LSP}^{2}
~,\ee
where $p$ is the final state momentum and we have assumed the $\tilde{l}_{R}$ are degenerate and that the $a$ term is dominated by the $\tau$ lepton contribution. Thus $T_{b}$ is given by 
\be{an6} T_{b} = \frac{1}{12} \left(\frac{m_{\tau}}{m_{LSP}}\right)^{2} m_{LSP} 
\approx 0.005 \left(\frac{50 \GeV}{m_{LSP}}\right)   \GeV     ~.\ee
Therefore, for the case of pure binos, $T_{b}$ will typically be less than about 5 MeV. We can therefore consider $T_{b} < T_{d}$ in the following. For example, for the case with $m_{LSP} \approx m_{\tilde{l}_{R}}$, which will give the tightest bound on $f_{B}$ for a given $T_{d}$ and $m_{LSP}$, we obtain, for $T_{R} > T_{d}$, 
\be{an3} f_{B} \leq f_{B\;c} \approx 0.8
\left(\frac{5 \times 10^{-11}}{\eta_{B}}\right) 
\left(\frac{1\GeV}{T_{d}}\right)^{2} 
\left(\frac{m_{\tilde{l}_{R}}}{100 \GeV}\right)^{3} 
~.\ee
Thus, with $m_{LSP} \approx m_{\tilde{l}_{R}} \approx 50 \GeV$, we find that $f_{B\;c} \approx 0.1 T_{d}^{-2} \GeV^{2}$. Therefore $T_{d} \lae 0.3\GeV$ will allow all values of $f_{B}$ to evade annihilations after B-ball decay. If $T_{d} \approx 1 \GeV$, the final
LSP density will correspond to the case where $f_{B} = f_{B\;c} \approx 0.1$. This shows that annihilations after B-ball decays can result in an LSP density compatible with MSSM dark matter constraints even if $f_{B} \approx 1$, whilst still having a similar number density of baryons and dark matter particles; for $f_{B\;c} \approx 0.1$, we would obtain $\sigma_{B} \approx 3$. 

             It is possible to reach only very broad conclusions at present, as the 
B-ball decay parameters $f_{B}$ and $T_{d}$ are unknown and, in addition, the results
depend on the reheating temperature after inflation. However, both $f_{B}$ and $T_{d}$ are, 
in principle, calculable in a given model: $f_{B}$ by solving the non-linear scalar field equations governing the formation of B-balls from the original Affleck-Dine condensate and $T_{d}$ by calculating the charge and decay rate of the B-balls accurately. $T_{d}$, which will depend explicitly on the reheating temperature after inflation, is the more model-dependent of the two. The reheating temperature can be estimated under the assumption that the baryon asymmetry originates from an Affleck-Dine condensate with CP violating phase of the order of 1, and, indeed, can be calculated given all the details of an inflation model, but $T_{R}$ is likely remain an important source of theoretical uncertainty in the B-ball decay scenario. However, it is quite possible that 
$T_{d}$ and $T_{R}$, by being sufficiently small and large relative to $T_{f}$ respectively, play no direct role in determining the final LSP density. 

               The B-ball decay scenario for MSSM dark matter is a natural alternative to 
the thermal relic LSP scenario, and has the considerable advantage of being able to explain the similarity of the baryon and dark matter densities. Should future experimental constraints on the parameters of the MSSM prove to be incompatible with thermal relic dark matter but consistent with B-ball decay dark matter for some set of B-ball parameters, it would strongly support the 
B-ball decay scenario. In particular, should the LSP mass be determined experimentally, the ratio of the number density baryons to dark matter would then be constrained by nucleosynthesis bounds on the baryon asymmetry. This would impose significant constraints on the reheating temperature and B-ball parameters, which, by comparing with the theoretical value of $f_{B}$, could even provide a "smoking gun" for the validity of the B-ball decay scenario, should annihilations happen to play no role in determining the present LSP density. 

\subsection*{Acknowledgements}   This work has been supported by the
 Academy of Finland and by a European Union Marie Curie Fellowship under EU contract number 
ERBFM-BICT960567.

\end{document}